\documentclass[aps,tightenlines,showpacs, showkeys, notitlepage]{revtex4-1}
\usepackage[latin9]{inputenc}
\usepackage{amsmath}
\usepackage{amssymb}

\makeatletter

\usepackage{dcolumn}
\usepackage{bm}

\makeatother

\begin{document}
\title{Maxwell-Modified Metric Affine Gravity}
\author{Oktay Cebecio\u{g}lu}
\email{ocebecioglu@kocaeli.edu.tr}

\author{Salih Kibaro\u{g}lu}
\email{salihkibaroglu@gmail.com}

\date{\today}
\begin{abstract}
We present a gauge formulation of the special affine algebra extended
to include an antisymmetric tensorial generator belonging to the tensor
representation of the special linear group. We then obtain a Maxwell
modified metric affine gravity action with a cosmological constant
term. We find the field equations of the theory and show that the
theory reduces to an Einstein-like equation for metric affine gravity
with the source added to the gravity equations with cosmological constant
$\mu$ contains linear contributions from the new gauge fields. The
reduction of the Maxwell metric affine gravity to Riemann-Cartan one
is discussed and the shear curvature tensor corresponding to the symmetric
part of the special linear connection is identified with the dark
energy. Furthermore, the new gauge fields interpreted as geometrical
inflaton vector fields which drive accelerated expansion.
\end{abstract}
\affiliation{Department of Physics, Kocaeli University, 41380 Kocaeli, Turkey}
\keywords{Dark energy, gauge field theory, inflation, metric-affine gravitation
theory}
\maketitle

\section{Introduction}

It is verified by the Solar System and cosmological tests that general
relativity(GR) provides an elegant and powerful formulation of gravitation
in terms of Riemannian geometry and forms our understanding of space-time
\citep{will2014}. Despite these successes, there are some reasons
to believe that general relativity is unable to explain some gravity
phenomena on both atomic and cosmological scales and should be either
modified or replaced by a new theory of gravity. Recently many papers
propose new types of dynamics to explain the dark energy phenomenon
\citep{copeland2001,Caldwell2005} as well as the dynamical role of
the cosmological constant \citep{frieman2008,Padmanabhan2009}.

It is also known that the cosmological term, usually associated with
the vacuum energy density, cannot be a valid theoretical explanation
for the accelerated expansion of the universe \citep{sola2013}. A
very different approach holds that cosmic acceleration is a manifestation
of new gravitational physics rather than dark energy, i.e., that it
involves a modification of the geometry as opposed to the stress-energy
tensor side of the Einstein equations \citep{frieman2008}. It is
important to point out that one can accommodate a generalized cosmological
constant in the gravity theory using extended algebras. A way of introducing
the generalized cosmological constant term using the Maxwell algebra
was presented in \citep{azgarraga2011} and even more, interestingly
it has been argued that by making use of the gauged Maxwell algebra
one can understand it as a source of an additional contribution to
the cosmological term in Einstein gravity.

Maxwell symmetry was introduced around forty years ago \citep{bacry1970,schrader1972},
but it is only recently that has attracted more attention after the
work of Soroka \citep{soroka2005} in 2005. The Maxwell symmetry is
the result of extending the Poincare symmetry by six additional tensorial
Abelian symmetry generators that make the four-momenta non-commutative.
Since then a variety of different Maxwell (super) symmetry algebras
with interesting geometric and physical properties have been constructed
and analyzed in the papers \citep{gomis2009,bonanos2009,bonanos2010A,bonanos2010B,bonanos2010C,durka2012,azcarraga2013A,azcarraga2014,fedoruk2013}.

By gauging Maxwell symmetries, one can define modified gravitational
theories that extend GR by including a generalized cosmological term
\citep{soroka2012,durka2011,cebecioglu2014,concha2015,concha2019A,concha2019B,kibaro=00011Flu2018,kibaro=00011Flu2020,cebecioglu2015,kibaro=00011Flu2019,azcaraga2013}.
Among these is the Maxwell extension of special affine symmetry and
its gauging which will be the focus of our attention in this paper.

In 1974 Yang \citep{yang1974} put forward a gauge theory of gravity
based on the affine group to construct a theory of (quantum) gravity
in the high energy limit \citep{hehl1989}. On the other hand, in
nature, there is no conservation law corresponding to the (special)
linear transformation and so the linear transformations must be dynamical,
i.e., spontaneously broken \citep{borisov1974}. Correspondingly,
the papers \citep{neeman1979B,neemann1988A,neemann1988B,lee1990,lopez-pinto1995}
suggested that the renormalizability and unitarity problems in quantum
gravity can be overcome by taking the affine group as the dynamical
group in a gauge theory of gravity with the help of generalized linear
connection \citep{hehl1995}. There exists a series of papers \citep{lord1978,hehl1978,hamamoto1978,lopez-pinto1995,julia1998,leclerc2006,sobreiro2011,tresguerres2000}
in which an affine gauge gravitation theory is considered.

Our paper has the following structure. In Section 2, following \citep{cebecioglu2015,kibaro=00011Flu2019},
we briefly review the Maxwell extension of the special-affine group,
$\mathcal{M}\mathcal{SA}\left(4,R\right)$. We also present the transformation
rules for the generalized coordinates (coset parameters) and the corresponding
differential realization of generators using the nonlinear realization
technique. In Section 3, we gauge the Maxwell special linear algebra
$\mathfrak{msa}\left(4,R\right)$ and find the gauge covariant quantities
to construct the gauge-invariant action. In Section 4, we introduced
$\mathcal{SL}\left(4,R\right)$ gauge covariant metric tensor in the
affine space needed for the metric affine gravity (MAG). In Section
5, we propose an action for Maxwell metric affine gravity by using
Euler or Gauss-Bonnet type topological action and derive the equations
of motion of corresponding action. We present our conclusions in Section
6.

\section{\i ntroduc\i ng the special-affine algebra and its maxwell extension}

We begin in this section by giving an overview of the Maxwell extension
of the special affine group. For a more complete description of the
details, the reader is referred to earlier works \citep{cebecioglu2015,kibaro=00011Flu2019}.
The special affine symmetry group $\mathcal{SA}\left(4,R\right)$
is given by the semi-direct product of the special linear group $\mathcal{SL}\left(4,R\right)$
and the translation group $\mathcal{T}\left(4\right)$ and are generated
by the fifteen special linear generators$\mathring{L}_{\,\,b}^{a}$
and by the four affine translation generators $P_{a}$, respectively.
The commutators of the generators obey the following algebra,

\begin{eqnarray}
\left[\mathring{L}_{\,\,b}^{a},\mathring{L}_{\,\,d}^{c}\right] & = & i\left(\delta_{\,\,b}^{c}\mathring{L}_{\,\,d}^{a}-\delta{}_{\,\,d}^{a}\mathring{L}{}_{\,\,b}^{c}\right),\nonumber \\
\left[\mathring{L}_{\,\,b}^{a},P_{c}\right] & = & -i\left(\delta_{\,\,c}^{a}P_{b}-\frac{1}{4}\delta_{\,\,b}^{a}P_{c}\right),\nonumber \\
\left[P_{a},P_{b}\right] & = & 0.
\end{eqnarray}
From this algebra, we can construct a group element by exponentiation,
\begin{equation}
g\left(x,\mathring{\omega}\right)=e^{ix^{a}\left(x\right)P_{a}}e^{i\mathring{\omega}_{\,a}^{b}\left(x\right)\mathring{L}_{\,b}^{a}},
\end{equation}
where $x^{a}\left(x\right),\,\mathring{\omega}{}_{\,\,a}^{b}\left(x\right)$
are the real parameters. The Maurer-Cartan (MC) 1-forms is defined
as $\Omega=-ig^{-1}dg$, here $g$ is the general element of the $\mathcal{SA}\left(4,R\right)$
group and the structure equation is given by

\begin{equation}
d\Omega+\frac{i}{2}\left[\Omega,\Omega\right]=0.\label{eq: MC eq}
\end{equation}
Thus, one can show that the MC 1-forms satisfy following equations,

\begin{eqnarray}
0 & = & d\Omega_{\,\,\,P}^{a}+\Omega_{\,\,\mathring{L}b}^{a}\wedge\Omega_{\,\,\,P}^{b}-\frac{1}{4}\Omega_{\,\,\mathring{L}}\wedge\Omega_{\,\,\,P}^{a},\nonumber \\
0 & = & d\Omega_{\,\,\mathring{L}b}^{a}+\Omega_{\,\,\mathring{L}c}^{a}\wedge\Omega_{\,\,\mathring{L}b}^{c},\label{eq: MC SA}
\end{eqnarray}
where the MC 1-forms $\Omega_{\,\,\,P}^{a}$ and $\Omega_{\,\,\mathring{L}b}^{a}$
correspond to translations and special-linear transformations in affine
space-time. By using the MC structure equations (\ref{eq: MC SA})
and making use of the methods presented in \citep{bonanos2009,bonanos2010C},
one can consider a Maxwell extension of the special affine algebra
by the antisymmetric generator $Z_{ab}$. The non-vanishing commutation
relations are 

\begin{eqnarray}
\left[\mathring{L}_{\,\,b}^{a},\mathring{L}_{\,\,d}^{c}\right] & = & i\left(\delta_{\,\,b}^{c}\mathring{L}_{\,\,d}^{a}-\delta{}_{\,\,d}^{a}\mathring{L}{}_{\,\,b}^{c}\right),\nonumber \\
\left[\mathring{L}_{\,\,b}^{a},P_{c}\right] & = & -i\left(\delta_{\,\,c}^{a}P_{b}-\frac{1}{4}\delta_{\,\,b}^{a}P_{c}\right),\label{eq:msa alg-1}
\end{eqnarray}
as well as the Maxwell extension

\begin{eqnarray}
\left[P_{a},P_{b}\right] & = & iZ_{ab},\nonumber \\
\left[\mathring{L}_{\,\,b}^{a},Z_{cd}\right] & = & i\left(\delta_{\,\,d}^{a}Z_{bc}-\delta_{\,\,c}^{a}Z_{bd}+\frac{1}{2}\delta_{\,\,b}^{a}Z_{cd}\right).\label{eq: msa alg-2}
\end{eqnarray}
The action of space-time symmetries on the fields, obtained as an
induced representation, is related to the nonlinear realization of
symmetries and are developed in reference \citep{coleman1969,callan1969,salam1969A,salam1969B}.
Therefore, when talking about these systems we find the coset construction
provides the appropriate language. We refer the reader to references
\citep{borisov1974,hamamoto1978,lopez-pinto1995,tresguerres2000}
for the derivation of the non-linear realisation with the $\mathcal{SA}\left(4,R\right)$
group. We now construct the non-linear realization corresponding to
the group $\mathcal{M}\mathcal{SA}\left(4,R\right)$ taking the $\mathcal{\mathcal{S}L}\left(4,R\right)$
to be a local symmetry. We therefore parametrize the coset elements
of the form 

\begin{equation}
K(x,\theta)=\frac{\mathcal{M}\mathcal{SA}}{SL}=e^{ix\cdot P}e^{i\theta\cdot Z},
\end{equation}
where $x^{a},\,\theta^{ab}$ are the coset parameters. Upon using
the definition of the transformation of the coset representative

\begin{equation}
g(a,\varepsilon,u)K(x,\theta)=K(x^{\prime},\theta^{\prime})h(\mathring{\omega}),
\end{equation}
where $h(\mathring{\omega})=e^{i\mathring{\omega}{}_{\,\,a}^{b}\mathring{L}{}_{\,\,b}^{a}}$
stands for the subgroup element, we find that the infinitesimal transformations
of the coset parameters are given by 
\begin{align}
\delta x^{a} & =a^{a}+u_{\,\,c}^{a}x^{c}-\frac{1}{4}ux^{a},\label{eq:var_x}\\
\delta\theta^{ab} & =\varepsilon^{ab}+u_{\,\,\,c}^{[a|}\theta^{c|b]}-\frac{1}{2}u\theta^{ab}-\frac{1}{4}a^{[a}x^{b]},\label{eq:var_teta}\\
\mathring{\omega}_{\,\,b}^{a} & =u_{\,\,b}^{a},
\end{align}
where the square brackets denote antisymmetrization of corresponding
indices and $a,\,\epsilon,\,u$ are the real parameters for affine
space-time translations, tensorial translations and special linear
symmetry transformation respectively. 

For the sake of completeness, we give the differential realization
of the symmetry generators

\begin{align}
P_{a} & =i\left(\partial_{a}-\frac{1}{2}x^{b}\partial_{ab}\right),\nonumber \\
Z_{ab} & =i\partial_{ab},\nonumber \\
\mathring{L}_{\,\,b}^{a} & =i\left(x^{a}\partial_{b}+2\theta^{ac}\partial_{bc}\right)-\frac{1}{4}\delta_{\,\,b}^{a}\left(x^{c}\partial_{c}+2\theta^{cd}\partial_{cd}\right),
\end{align}
where $\theta$ derivative is defined by $\partial_{ab}\theta^{cd}=\frac{1}{2}\left(\delta_{\,\,a}^{c}\delta_{\,\,b}^{d}-\delta_{\,\,b}^{c}\delta_{\,\,a}^{d}\right)$.
It is an easy task to check that the generators satisfy the algebra
Eqs.(\ref{eq:msa alg-1}) and (\ref{eq: msa alg-2}).

\section{Gauging the Maxwell-spec\i al-affine algebra}

Let us construct a gauge theory for the Maxwell special affine algebra
$\mathfrak{msa}\left(4,R\right)$. For this purpose, we follow the
same methods given in \citep{azcarraga2014,cebecioglu2014,cebecioglu2015}.
The gauge field is a $\mathfrak{msa}\left(4,R\right)$ algebra valued
one-form 
\begin{equation}
\mathcal{A}=e^{a}P_{a}+B^{ab}Z_{ab}+\mathring{\omega}_{\,\,a}^{b}\mathring{L}_{\,\,b}^{a}.\label{eq:A-form}
\end{equation}
An infinitesimal gauge parameter is 

\begin{equation}
\zeta\left(x\right)=y^{a}\left(x\right)P_{a}+\varphi^{ab}(x)Z_{ab}+\lambda{}_{\,\,a}^{b}(x)\mathring{L}{}_{\,\,b}^{a},
\end{equation}
where $y^{a}\left(x\right)$, $\varphi^{ab}(x)$, and $\lambda{}_{\,\,a}^{b}(x)$
are the infinitesimal parameters corresponding to the affine translation,
tensorial and special linear transformations respectively.

The gauge transformation are given by

\begin{equation}
\delta\mathcal{A}=-d\zeta-i\left[\mathcal{A},\zeta\right],\label{delta_amu}
\end{equation}
evaluating (\ref{eq:A-form}), we get 
\begin{align}
\delta e^{a} & =-dy^{a}-\mathring{\omega}{}_{\,\,b}^{a}y^{b}+\frac{1}{4}\mathring{\omega}y^{a}+\lambda_{\,\,b}^{a}e^{b}-\frac{1}{4}\lambda e^{a}=-\mathfrak{D}y^{a}+\lambda_{\,\,b}^{a}e^{b}-\frac{1}{4}\lambda e^{a},\label{eq:vierbein_geuge_trans}\\
\delta B^{ab} & =-d\varphi^{ab}-\mathring{\omega}_{\,\,\,c}^{[a|}\varphi^{cb]}+\frac{1}{2}\mathring{\omega}\varphi^{ab}+\lambda_{\,\,\,c}^{[a|}B^{cb]}-\frac{1}{2}\lambda B^{ab}+\frac{1}{2}e^{[a}y^{b]}\\
 & =-\mathfrak{D}y^{a}+\lambda_{\,\,\,c}^{[a|}B^{cb]}-\frac{1}{2}\lambda B^{ab}+\frac{1}{2}e^{[a}y^{b]},\\
\delta\mathring{\omega}_{\,\,b}^{a} & =-d\lambda_{\,\,b}^{a}-\mathring{\omega}_{\,\,c}^{a}\lambda_{\,\,b}^{c}+\mathring{\omega}_{\,\,b}^{c}\lambda_{\,\,c}^{a}=-\mathcal{\mathfrak{D}}\lambda_{\,\,b}^{a},
\end{align}
where the $\mathcal{SL}\left(4,R\right)$ valued exterior covariant
derivative $\mathfrak{D}$ of a tensor density $\Phi$ of affine weight
$w$ contains $\left\{ wTr\left(\mathring{\omega_{b}^{a}}\right)\Phi\right\} $,
eg. 
\begin{equation}
\left(\mathcal{\mathfrak{D}}\Phi\right)_{b}^{a}=\left[\delta_{b}^{a}d+\mathring{\omega_{b}^{a}}+w\left(\Phi\right)Tr\left(\mathring{\omega}_{b}^{a}\right)\right]\Phi.
\end{equation}
From transformation rules, we immediately infer that 1-forms $e^{a}$,
$B^{ab}$, and $\mathring{\omega}{}_{\,\,a}^{b}$ have the following
affine scaling weights $-1/4$, $-1/2$, and $0$ respectively.

Now, acting the exterior covariant derivative on $\mathcal{A}$ we
obtain the curvature $\digamma$satisfying the structure equation
and the Bianchi identity
\begin{equation}
\digamma=d\mathcal{A}+\frac{i}{2}\left[\mathcal{A},\mathcal{A}\right],\label{eq:str}
\end{equation}

\begin{equation}
d\digamma+i\left[\mathcal{A},\digamma\right]=0,\label{eq:bianchi}
\end{equation}
where $d$ is the exterior differential. Upon expressing the curvature
form $\digamma$ as
\begin{equation}
\mathcal{\digamma}=\mathcal{F}^{a}P_{a}+\mathcal{F}^{ab}Z_{ab}+\mathcal{R}_{\,\,b}^{a}\mathring{L}{}_{\,\,a}^{b},\label{f1}
\end{equation}
 the structure Eq.(\ref{eq:str}) becomes

\begin{eqnarray}
\mathcal{F}^{a} & = & de^{a}+\mathring{\omega}_{\,\,b}^{a}\wedge e^{b}-\frac{1}{4}\mathring{\omega}\wedge e^{a}=\mathcal{\mathfrak{D}}e^{a},\\
\mathcal{F}^{ab} & = & dB^{ab}+\mathring{\omega}_{\,\,\,c}^{[a|}\wedge B^{c|b]}-\frac{1}{2}\mathring{\omega}\wedge B^{ab}-\frac{1}{2}e^{a}\wedge e^{b},\label{eq: curv_Fab}\\
 & = & \mathcal{\mathfrak{D}}B^{ab}-\frac{1}{2}e^{a}\wedge e^{b},\\
\mathcal{R}_{\,\,b}^{a} & = & d\mathring{\omega}_{\,\,b}^{a}+\mathring{\omega}_{\,\,c}^{a}\wedge\mathring{\omega}_{\,\,b}^{c}=\mathcal{\mathfrak{D}}\mathring{\omega}_{\,\,b}^{a}.
\end{eqnarray}
Thus the curvature forms corresponding to the various generators of
the algebra are $\left(\mathcal{F}^{a},\mathcal{F}^{ab},\mathcal{R}_{\,\,b}^{a}\right)$,
and they represent the torsion, the field strength associated with
the $B^{ab}$ field and the non-Riemannian affine curvature form,
respectively. One concludes that the affine curvature $\mathcal{R}_{\,\,b}^{a}$
and the torsion $\mathcal{F}^{a}$ are given by the exterior covariant
derivatives of the affine connection and vierbein respectively. On
the other hand, the curvature 2-form $\mathcal{F}^{ab}$ coming from
Maxwell extension is not given by the exterior covariant derivative
of the corresponding gauge field. The extra term in $\mathcal{F}^{ab}$
represents the curvature of the local tensor space. This contribution
is present because the commutator of two infinitesimal affine translations
equals to an element of the tensor space. Moreover, from the Bianchi
identity Eq.(\ref{eq:bianchi}), we get the following equations

\begin{eqnarray}
\mathcal{\mathfrak{D}}\mathcal{F}^{ab} & = & \mathcal{R}_{\,\,\,c}^{[a|}\wedge B^{c|b]}-\frac{1}{2}\mathcal{R}\wedge B^{ab}-\frac{1}{2}\mathcal{F}^{[a}\wedge e^{b]},\nonumber \\
\mathcal{\mathfrak{D}}\mathcal{F}^{a} & = & \mathcal{R}_{\,\,b}^{a}\wedge e^{b}-\frac{1}{4}\mathcal{R}\wedge e^{a},\nonumber \\
\mathcal{\mathfrak{D}}\mathcal{R}_{\,\,b}^{a} & = & 0.\label{eq:Bianchi}
\end{eqnarray}
Under infinitesimal gauge transformations with parameter $\zeta$,
the curvature 2-form $\digamma$ transform as 

\begin{equation}
\delta\mathcal{\digamma}=i\left[\zeta,\mathcal{\digamma}\right],
\end{equation}
and hence one gets
\begin{eqnarray}
\delta\mathcal{F}^{a} & = & -\mathcal{R}_{\,\,b}^{a}y^{b}+\frac{1}{4}\mathcal{R}y^{a}+\lambda_{\,\,b}^{a}\mathcal{F}^{b}-\frac{1}{4}\lambda\mathcal{F}^{a},\\
\delta\mathcal{F}^{ab} & = & -\mathcal{R}_{\,\,\,c}^{[a|}\varphi^{c|b]}+\frac{1}{2}\mathcal{R}\varphi^{ab}+\lambda_{\,\,\,c}^{[a|}F^{c|b]}-\frac{1}{2}\lambda\mathcal{F}^{ab}+\frac{1}{2}\mathcal{F}^{[a}y^{b]},\label{eq:fab}\\
\delta\mathcal{R}_{\,\,b}^{a} & = & \lambda_{\,\,c}^{a}\mathcal{R}_{\,\,b}^{c}-\lambda_{\,\,b}^{c}\mathcal{R}_{\,\,c}^{a}.
\end{eqnarray}
Again from these transformation rules, one observes that curvature
2-form $\mathcal{F}^{a}$, $\mathcal{F}^{ab}$, and $\mathcal{R}_{\,\,b}^{a}$
have the following affine scaling weights $-1/4$, $-1/2$, and $0$
respectively and they will be useful for constructing invariant Lagrangian
densities.

\section{construct\i on of the metric for the affine space}

Using the definition of the local metric,

\begin{equation}
g^{ab}\left(x\right)=e^{a}\otimes e^{b},
\end{equation}
one deduces $\mathcal{\mathcal{S}L}\left(4,R\right)$ gauge variation
of the metric tensor with the help of Eq.(\ref{eq:vierbein_geuge_trans})
by omitting diffeomorphism part

\begin{eqnarray}
\delta_{\lambda}g^{ab} & = & \left(\lambda_{\,\,c}^{a}e^{c}-\frac{1}{4}\lambda e^{a}\right)\otimes e^{b}+e^{a}\otimes\left(\lambda_{\,\,c}^{b}e^{c}-\frac{1}{4}\lambda e^{b}\right)=\lambda_{\,\,\,c}^{(a}g^{cb)}-\frac{1}{2}\lambda g^{ab},
\end{eqnarray}
where round brackets denote symmetrization. Similarly from the definition
of the Kronecker delta tensor

\begin{equation}
\delta_{\,\,b}^{a}=e^{a}\otimes e_{b},
\end{equation}
one can obtain the $\mathcal{\mathcal{S}L}\left(4,R\right)$ gauge
variation of $e_{a}$ as

\begin{equation}
\delta_{\lambda}e_{a}=-\lambda_{\,\,a}^{b}e_{b}+\frac{1}{4}\lambda e_{a},
\end{equation}
and the last equation implies
\begin{eqnarray}
\delta_{\lambda}g_{ab} & = & -\lambda_{\,\,(a}^{c}g_{cb)}+\frac{1}{2}\lambda g_{ab}.
\end{eqnarray}
With the use of vierbein and local metric, the $\mathcal{\mathcal{S}L}\left(4,R\right)$
gauge variation of the coordinate metric becomes

\begin{eqnarray}
\delta_{\lambda}g_{\mu\nu}\left(x\right) & = & \left(-\lambda_{\,\,(a}^{c}g_{cb)}+\frac{1}{2}\lambda g_{ab}\right)e_{\mu}^{a}e_{\nu}^{b}+g_{ab}\left(\lambda_{c}^{a}e_{\mu}^{c}-\frac{1}{4}\lambda e_{\mu}^{a}\right)e_{\nu}^{b}+g_{ab}e_{\mu}^{a}\left(\lambda_{\,\,c}^{b}e_{\nu}^{c}-\frac{1}{4}\lambda e_{\nu}^{b}\right)=0.
\end{eqnarray}
Moreover, the gauge variation of determinant of the vierbein is

\begin{equation}
\delta_{\lambda}e=\frac{1}{2}eg^{\mu\nu}\delta_{\lambda}g_{\mu\nu}=0.
\end{equation}
Defining the fully antisymmetric tensor $\eta_{abcd}$ by

\begin{equation}
\eta_{abcd}=e\varepsilon_{abcd},\label{eq:levi_civita}
\end{equation}
where $\varepsilon_{abcd}$ is the Levi-Civita symbol, its variation
under local $\mathcal{\mathcal{S}L}\left(4,R\right)$ transformation
becomes
\begin{eqnarray}
\delta_{\lambda}\eta_{abcd} & = & -\lambda_{\,\,a}^{e}\eta_{ebcd}-\lambda_{\,\,b}^{e}\eta_{aecd}-\lambda_{\,\,c}^{e}\eta_{abed}-\lambda_{\,\,d}^{e}\eta_{abce}+\lambda\eta_{abcd}=0,
\end{eqnarray}
and has affine scaling weight $1$.

Having defined the local metric for the affine space-time, the metricity
is obtained by taking the covariant derivative of the local metric,
i.e., $Q^{ab}=\mathcal{D}g^{ab}$ and its explicit form follows

\begin{eqnarray}
Q^{ab} & = & \mathcal{\mathfrak{D}}g^{ab}=dg^{ab}+\mathring{\omega}_{\,\,c}^{(a}g^{cb)}-\frac{1}{2}\mathring{\omega}g^{ab},\nonumber \\
Q_{ab} & = & \mathcal{\mathfrak{D}}g_{ab}=dg_{ab}-\mathring{\omega}_{\,\,(a}^{c}g_{cb)}+\frac{1}{2}\mathring{\omega}g_{ab}.
\end{eqnarray}
This in turn leads to the covariant derivative of the metricity

\begin{eqnarray*}
\mathcal{\mathfrak{D}}Q^{ab} & = & \mathcal{R}_{\,\,\,c}^{(a}g^{cb)}-\frac{1}{2}\mathcal{R}g^{ab}.
\end{eqnarray*}
Likewise,

\begin{equation}
\mathcal{\mathfrak{D}}Q_{ab}=-\mathcal{R}_{\,\,(a}^{c}g_{cb)}+\frac{1}{2}\mathcal{R}g_{ab}.
\end{equation}

\section{maxwell-modified mag field equations}

One way of constructing the action is to begin from the covariant
quantities with manifest geometric meanings. To prescribe the dynamics
of the gauge fields, we have to introduce an action, invariant under
local $\mathcal{SL}\left(4,R\right)$ transformation. We need then
curvatures $\mathcal{R}_{\,\,b}^{a}$, $\mathcal{F}^{ab}$ and the
metric $g^{ab}$ obtained in the last section. We start with following
topological action,

\begin{equation}
S=\frac{1}{2\chi}\int\mathcal{J}\wedge*\mathcal{J}=\frac{1}{4\chi}\int\eta_{abcd}\mathcal{J}^{ab}\wedge\mathcal{J}^{cd},\label{eq:lagrangian_euler}
\end{equation}
known as Euler or Gauss-Bonnet type action, where $\chi=8\pi G/c^{4}$
is the Einstein's constant, $\left(*\right)$ is the Hodge dual and
$\eta_{abcd}$ is defined by Eq.(\ref{eq:levi_civita}). Contracting
$\mathcal{R}_{\,\,b}^{a}$ with $g^{ab}$, we can form curvature 2-form
$\mathcal{R}^{ab}=\mathcal{R}_{\,\,c}^{a}g^{cb}$ and it's gauge transformation
is given by

\begin{eqnarray}
\delta\mathcal{R}^{ab} & = & \lambda_{\,\,c}^{a}\mathcal{R}^{cb}+\lambda_{\,\,c}^{b}\mathcal{R}^{ac}-\frac{1}{2}\lambda\mathcal{R}^{ab}.\label{eq:gauge_trans_Rab_shift}
\end{eqnarray}
It has the same form as Eq.(\ref{eq:fab}) when the diffeomorphism
part omitted. So, one can introduce a shifted curvature 2-form,

\begin{equation}
\mathcal{J}^{ab}=\mathcal{R}^{ab}-\mu\mathcal{F}^{ab},\label{eq:shifted_curvature}
\end{equation}
where $\mu$ is a dimensionful constant. Its gauge transformation
becomes

\begin{equation}
\delta\mathcal{J}^{ab}=\lambda_{\,\,c}^{a}\mathcal{J}^{cb}+\lambda_{\,\,c}^{b}\mathcal{J}^{ac}-\frac{1}{2}\lambda\mathcal{J}^{ab}.\label{eq:var_Jab}
\end{equation}
The gauge transformation of $*\mathcal{J}_{ab}=\frac{1}{2}\eta_{abcd}\mathcal{J}^{cd}$
has the following form
\begin{eqnarray}
\delta*\mathcal{J}_{ab} & = & \frac{1}{2}\left(\lambda_{\,\,c}^{e}\eta_{abed}+\lambda_{\,\,d}^{e}\eta_{abce}-\frac{1}{2}\lambda\eta_{abcd}\right)\mathcal{J}^{cd},\label{eq:var_J_star_1}
\end{eqnarray}
the term in the parentheses can be written another form after re-indeksing
the labels as
\begin{eqnarray}
\lambda_{\,\,c}^{e}\eta_{abed}+\lambda_{\,\,d}^{e}\eta_{abce}-\frac{1}{2}\lambda\eta_{abcd} & = & -\lambda_{\,\,a}^{e}\eta_{ebcd}-\lambda_{\,\,b}^{e}\eta_{aecd}+\frac{1}{2}\lambda\eta_{abcd},\label{eq:var_eta_1}
\end{eqnarray}
then variation of the Hodge dual of $J$ becomes

\begin{eqnarray}
\delta*\mathcal{J} & = & -\lambda_{\,\,a}^{e}*\mathcal{J}_{eb}-\lambda_{\,\,b}^{e}*\mathcal{J}_{ae}+\frac{1}{2}\lambda*\mathcal{J}_{ab}.\label{eq:var_Jab_hodge}
\end{eqnarray}
Invariance of the action under gauge transformation can be checked
easily with the help of Eqs.(\ref{eq:var_Jab}) and (\ref{eq:var_Jab_hodge}).
By construction, the action is automatically invariant under diffeomorphism
and has affine scaling weight zero.

It remains, of course, to find the field equations for the gauge fields.
The variation of the action (\ref{eq:lagrangian_euler}) with respect
to gauge fields $\mathring{\omega}{}_{\,\,a}^{b}$, $e^{a}\left(x\right)$,
$B^{ab}\left(x\right)$ and the metric $g^{ab}\left(x\right)$ lead
to the following equations:

\begin{eqnarray}
\mathcal{\mathfrak{D}}\left(g^{ac}*\mathcal{J}_{bc}\right)-2\mu B^{ac}\wedge*\mathcal{J}_{bc} & = & 0,\\
e^{b}\wedge*\mathcal{J}_{ab} & = & 0,\label{eq:var_S_e}\\
\mathfrak{D}\left(*\mathcal{J}_{ab}\right) & = & 0,\\
\mathcal{R}_{\,\,(a}^{c}\wedge*\mathcal{J}_{cb)}-\frac{1}{2}g_{ab}\mathcal{J}^{cd}\wedge*\mathcal{J}_{cd} & = & 0.
\end{eqnarray}
It is important to note that these equations of motion transform as
$\mathcal{SL}\left(4,R\right)$-tensors. Here, the first and second
equations represent the generalizations of the torsion equation, and
the Einstein equation. The third equation arises from the Maxwell
symmetry and the last equation is the generalized version of the equation
of motion for the metric tensor given in \citep{julia1998}.

In order to write Eq.(\ref{eq:var_S_e}) in the more conventional
form, one switches from tangent indices to coordinate indices,

\begin{eqnarray}
*\mathcal{J}_{cd}\wedge e^{c} & = & \frac{1}{2}e\varepsilon_{abcd}\mathcal{J}^{ab}\wedge e^{c}\nonumber \\
 & = & \frac{1}{4}e\varepsilon_{abcd}\mathcal{J}_{\mu\nu}^{ab}e_{\alpha}^{c}dx^{\mu}\wedge dx^{\nu}\wedge dx^{\alpha},
\end{eqnarray}
multiplying this from right $dx^{\beta}$, we get following equation

\begin{equation}
\mathcal{J}_{\,\,\nu}^{\mu}-\frac{1}{2}\delta_{\,\,\nu}^{\mu}\mathcal{J}=0,
\end{equation}
which can also be expressed as

\begin{equation}
\mathcal{R}_{\,\,\nu}^{\mu}-\frac{1}{2}\delta_{\,\,\nu}^{\mu}\mathcal{R}=\mu\left(\mathcal{F}_{\,\,\nu}^{\mu}-\frac{1}{2}\delta_{\,\,\nu}^{\mu}\mathcal{F}\right).\label{eq:field_eq_RF}
\end{equation}
This has resemblance to the usual Einstein's field equation. However,
the curvature tensor $\mathcal{R}_{\,\,\nu}^{\mu}$ and $\mathcal{F}_{\,\,\nu}^{\mu}$
may not necessarily be symmetric. $\mathcal{F}_{\,\,\nu}^{\mu}$ acts
as sources in the field equation of gravity. This equation can be
written in a more familiar form by going from differential form to
space-time tensors as

\begin{equation}
\frac{1}{2}\mathcal{F}_{\,\,\,\,\rho\sigma}^{\mu\nu}dx^{\rho}\wedge dx^{\sigma}=\frac{1}{2}\left(e_{a}^{\mu}e_{b}^{\nu}\mathcal{\mathcal{\mathfrak{D}}}_{[\rho}B_{\sigma]}^{ab}-\frac{1}{2}\delta_{\rho}^{\mu}\delta_{\sigma}^{\nu}+\frac{1}{2}\delta_{\sigma}^{\mu}\delta_{\rho}^{\nu}\right)dx^{\rho}\wedge dx^{\sigma},
\end{equation}
so we get explicit form of $\mathcal{F}_{\,\,\,\,\rho\sigma}^{\mu\nu}$,

\begin{equation}
\mathcal{F}_{\,\,\,\,\rho\sigma}^{\mu\nu}=e_{a}^{\mu}e_{b}^{\nu}\mathcal{\mathcal{\mathfrak{D}}}_{[\rho}B_{\sigma]}^{ab}-\frac{1}{2}\delta_{\rho}^{\mu}\delta_{\sigma}^{\nu}+\frac{1}{2}\delta_{\sigma}^{\mu}\delta_{\rho}^{\nu},
\end{equation}
then $F_{\,\,\rho}^{\mu}$ and $F$ can be extracted respectively
as,

\begin{eqnarray}
\mathcal{F}_{\,\,\rho}^{\mu} & = & \mathcal{F}_{\,\,\,\,\rho\nu}^{\mu\nu}=e_{a}^{\mu}e_{b}^{\nu}\mathcal{\mathcal{\mathfrak{D}}}_{[\rho}B_{\nu]}^{ab}-\frac{3}{2}\delta_{\rho}^{\mu},
\end{eqnarray}

\begin{equation}
\mathcal{F}=\mathcal{F}_{\,\,\mu}^{\mu}=e_{a}^{\mu}e_{b}^{\nu}\mathcal{\mathcal{\mathfrak{D}}}_{[\mu}B_{\nu]}^{ab}-6.
\end{equation}
Thanks to the last three equations, we can re-expressed the right
hand side of Eq.(\ref{eq:field_eq_RF}),

\begin{equation}
\mathcal{F}_{\,\,\nu}^{\mu}-\frac{1}{2}\delta_{\,\,\nu}^{\mu}\mathcal{F}=e_{a}^{\mu}e_{b}^{\rho}\mathcal{\mathfrak{D}}_{[\nu}B_{\rho]}^{ab}-\frac{1}{2}\delta_{\nu}^{\mu}e_{a}^{\rho}e_{b}^{\sigma}\mathcal{\mathfrak{D}}_{[\rho}B_{\sigma]}^{ab}+\frac{3}{2}\delta_{\nu}^{\mu},
\end{equation}
so the Eq.(\ref{eq:field_eq_RF}) takes the following form,

\begin{equation}
\mathcal{R}_{\,\,\nu}^{\mu}-\frac{1}{2}\delta_{\nu}^{\mu}\mathcal{R}-\frac{3}{2}\mu\delta_{\nu}^{\mu}=\mu\left(e_{a}^{\mu}e_{b}^{\rho}\mathcal{\mathfrak{D}}_{[\nu}B_{\rho]}^{ab}-\delta_{\nu}^{\mu}e_{a}^{\rho}e_{b}^{\sigma}\mathcal{\mathfrak{D}}_{\rho}B_{\sigma}^{ab}\right),\label{eq:field_eq_RF-1}
\end{equation}
where $\frac{1}{2}\delta_{\nu}^{\mu}e_{a}^{\rho}e_{b}^{\sigma}\mathcal{\mathfrak{D}}_{[\rho}B_{\sigma]}^{ab}=\delta_{\nu}^{\mu}e_{a}^{\rho}e_{b}^{\sigma}\mathcal{\mathfrak{D}}_{\rho}B_{\sigma}^{ab}$.

We see that the source added to the gravity equations with cosmological
constant $\mu$ contains linear contributions from the new gauge fields.
The second term on the right-hand side of (\ref{eq:field_eq_RF-1})
provides a field-dependent modification of the cosmological constant
at the left-hand side of the equation\citep{azcaraga2013}.

To the decomposition above there corresponds a splitting of the connection
1-form into its Riemannian and non-Riemannian parts $\omega_{\,\,b}^{a}$
and , $v_{\,\,b}^{a}$, respectively, as

\begin{equation}
\mathring{\omega}{}_{\,\,a}^{b}=\omega_{\,\,b}^{a}+v_{\,\,b}^{a},
\end{equation}
where $\omega_{\,\,b}^{a}$ is antisymmetric Lorentz connection and
$v_{\,\,b}^{a}$ is symmetric shear connection. In terms of these
forms Eq.(\ref{eq:field_eq_RF-1}) becomes,

\begin{eqnarray}
\mathcal{R}_{\,\,\nu}^{\mu}-\frac{1}{2}\delta_{\nu}^{\mu}\mathcal{R}-\frac{3}{2}\mu\delta_{\nu}^{\mu} & = & \mu e_{a}^{\mu}e_{b}^{\rho}\left(D_{[\nu}B_{\rho]}^{ab}+v_{[\nu\,c}^{[a}\wedge B_{\rho]}^{cb]}-\frac{1}{2}v_{[\nu}B_{\rho]}^{ab}\right)\nonumber \\
 &  & -\mu\delta_{\nu}^{\mu}e_{a}^{\rho}e_{b}^{\sigma}\left(D_{\rho}B_{\sigma}^{ab}+v_{\rho\,c}^{[a}\wedge B_{\sigma}^{cb]}-\frac{1}{2}v_{\rho}B_{\sigma}^{ab}\right),
\end{eqnarray}
where $D$ is the Lorentz exterior covariant derivative. We see that
this is simply Einstein's equation for metric affine gravity with
a cosmological constant term. It is then sensible to identify the
expression in the curly bracket as the source of the gravitational
field. Note also that, if the affine curvature tensor is decomposed
into the Riemannian and shear strength tensor parts, it can end up
in a Riemann-Cartan theory with two extra fields, i.e., translational
connection and the symmetric sector of the $\mathcal{SL}\left(4,R\right)$-connection
both migrate to the stress-energy sector. This means that shear strength
tensor, considered to be an intrinsic property of metric affine space-time,
may measure the energy content of the universe, i.e., dark energy.

\section{Conclusion}

In conclusion, we have used coset formalism to determine geometrically
the Maxwell algebras for gauge-special affine gravity. For this purpose,
we have introduced a $\mathcal{SL}\left(4,R\right)$ connection over
a $(x^{a},\theta^{ab})=\left(4,6\right)$-dimensional tensor extended
space obtained by extending a metric affine space-time with six tensor
coordinates. After gauging the Maxwell special affine group $\mathcal{MSA}\left(4,R\right)$,
we propose a locally $\mathcal{SL}\left(4,R\right)$ invariant action
for the Maxwell extended MAG with the help of topological Euler or
Gauss-Bonnet type action. It is found that the Maxwell extension modifies
the results of the metric affine gravity not only by changing the
numerical coefficient of the cosmological term but also the new abelian
gauge fields $B_{\mu}^{ab}(x)$ terms already present in the latter
theory and they can be interpreted as geometrical inflaton vector
fields \citep{azcaraga2013} which drive accelerated expansion.
\begin{acknowledgments}
This work was supported by the Scientific and Technological Research
Council of Turkey (TÜB\.{I}TAK) Research project No. 118F364.
\end{acknowledgments}

\end{document}